\documentclass[a4paper,11pt]{article}
\usepackage{pos}

\title{A combined explanation of the $B$-meson decay anomalies with a single vector leptoquark}
\ShortTitle{A combined explanation of the $B$ anomalies with a single vector LQ}
\author*[a]{Jonathan Kriewald}
\author[b]{Chandan Hati}
\author[a]{Jean Orloff}
\author[a]{Ana M. Teixeira}
\affiliation[a]{Laboratoire de Physique de Clermont (LPC), CNRS/IN2P3 UMR 6533\\ 
Campus Universitaire des Cézeaux, Université Clermont-Auvergne\\
4 Avenue Blaise Pascal, TSA 60026, F-63178 Aubière Cedex, France\\}
\affiliation[b]{Physik Department T70, Technische Universit\"at M\"unchen,\\
James-Franck-Stra{\ss}e 1, D-85748 Garching, Germany}

\emailAdd{jonathan.kriewald@clermont.in2p3.fr}
\emailAdd{c.hati@tum.de}
\emailAdd{orloff@in2p3.fr}
\emailAdd{ana.teixeira@clermont.in2p3.fr}

\abstract{Motivated by the recent experimental progress on the $R_{D^{(\ast)}}$ and
$R_{K^{(\ast)}}$ anomalies in $B$-decays, we consider an extension
of the Standard Model by a single vector leptoquark field. 
We study how one can achieve the required lepton flavour non-universality,
starting from a priori universal gauge couplings.  
While the unitary coupling flavour structure, induced by the mass misalignment of quarks and leptons after $SU(2)_L$ breaking, does not allow to comply with stringent bounds from lepton flavour violating processes, 
we find that effectively non-unitary couplings, due to mixings with heavy vector-like fermions, hold the key to
simultaneously address the $R_{K^{(\ast)}}$ and $R_{D^{(\ast)}}$
anomalies.  
Furthermore, in the near future, the expected progress in the sensitivity on charged lepton flavour violating observables should allow to probe a large region of the preferred parameter space in this class of vector leptoquark models.}

\FullConference{%
  40th International Conference on High Energy physics - ICHEP2020\\
  July 28 - August 6, 2020\\
  Prague, Czech Republic (virtual meeting)
}

\begin{document}
\maketitle
\section{Introduction and Motivation}
Despite the discovery of the Higgs boson at the LHC, direct signals for new particles have so far eluded experimental observation.
However, in recent years, indirect hints of new physics (NP) in flavour observables have begun to manifest: in particular, precise measurements of semi-leptonic charged and neutral current decays of $B$-mesons appear to show deviations from the SM predictions. 
The ratios $\pmb{R_{D^{(\ast)}}} = \mathrm{BR}(B\to D^{(\ast)}\tau\nu)/\mathrm{BR}(B\to D^{(\ast)}\ell\nu)$~\cite{Belle:2019rba} and $\pmb{R_{K^{(\ast)}}} = \mathrm{BR}(B\to K^{(\ast)}\mu\mu)/\mathrm{BR}(B\to K^{(\ast)}ee)$~\cite{Aaij:2019wad,Aaij:2017vbb} are, up to kinematical factors, expected to be unity in the SM.
Both processes currently exhibit a discrepancy of around $\sim 3\:\sigma$, hinting indirectly at the presence of $\mathrm{TeV}$-scale NP with lepton flavour universality violating (LFUV) interactions.
Moreover, angular observables in $B\to K^\ast \mu\mu$ decays also exhibit a pattern of deviations with local discrepancies around $3\:\sigma$ as well~\cite{Aaij:2015oid, Aaij:2020nrf}.
Due to their ability to generate sizeable contributions to semileptonic flavour changing neutral currents (FCNC) at tree-level, leptoquark (LQ) models are currently under extensive scrutiny.

\section{A vector leptoquark model}
\begin{figure}
    \centering
    \mbox{\hspace{-8mm}
    \includegraphics[width=0.44\textwidth]{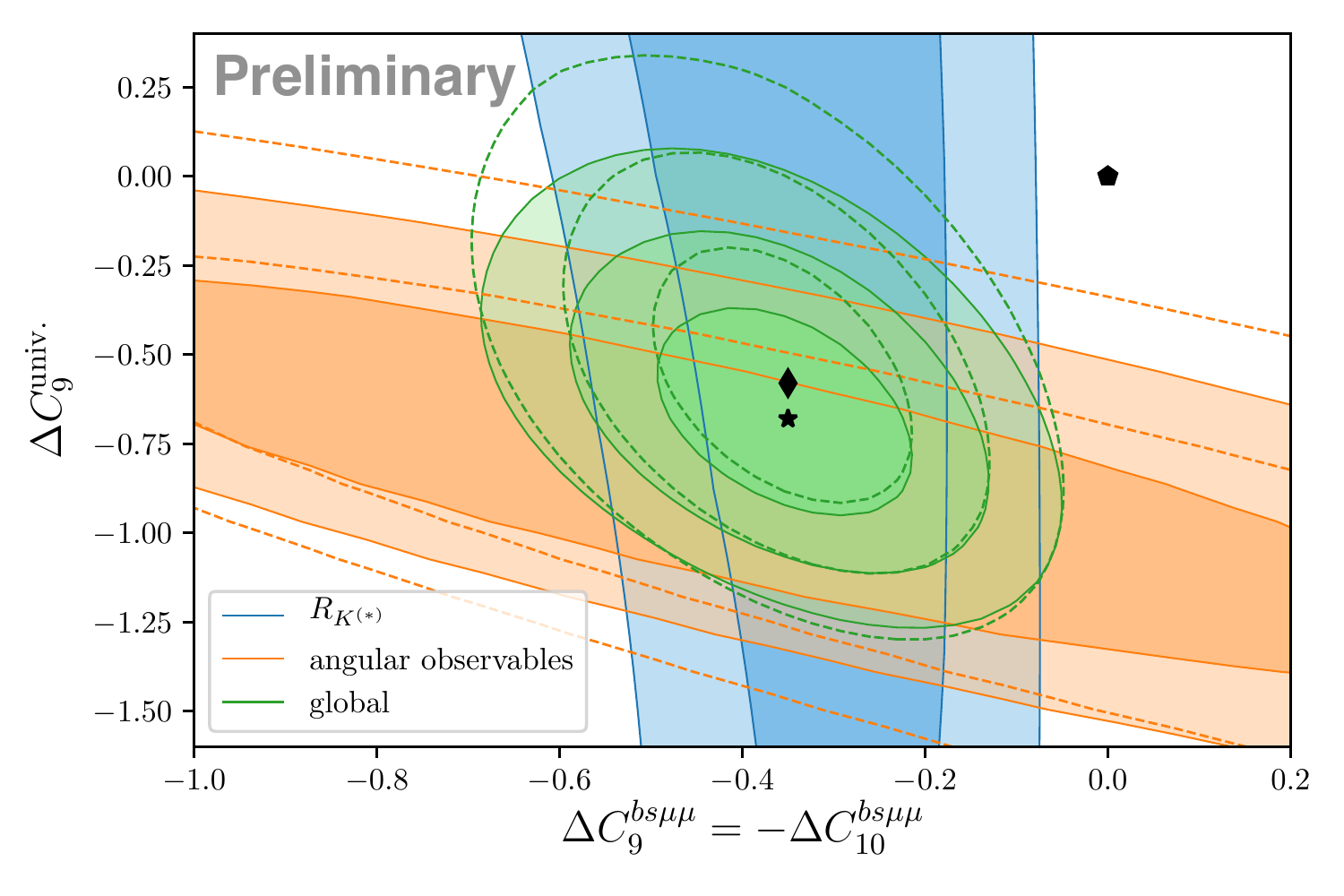}
    \includegraphics[width=0.48\textwidth]{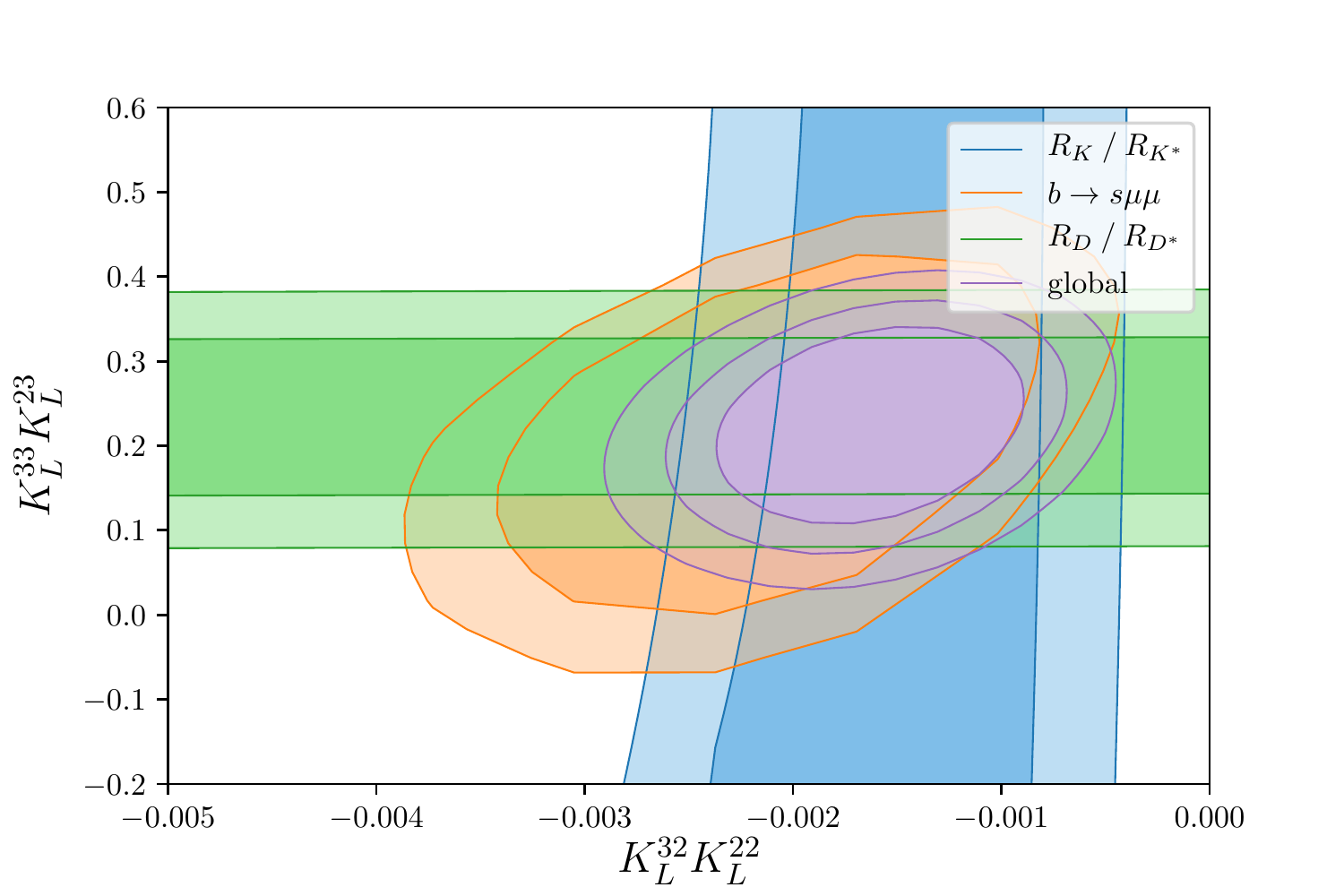}
    }
    \caption{{\bf Left:} Likelihood contours resulting from a fit to all available $b\to s\mu\mu$ data in the plane of ($V-A$) NP contributions vs. an additional {\it universal} contribution. The filled contours include the recent LHCb measurement~\cite{Aaij:2020nrf}, while the likelihood contours depicted by dashed lines exclude this data. The pentagon denotes the SM prediction, while the star (diamond) denotes the new (old) best fit point. {\bf Right:} Likelihood contours resulting from a fit of the dominant LQ coupling products to both $b\to s\ell\ell$ and $b\to c\tau\nu$ data.}
    \label{fig:plots1}
\end{figure}

Notably, the $SU(2)_L$-singlet vector LQ $V_1$ (transforming as $(\mathbf{3}, \mathbf{1}, 2/3)$ under the SM gauge group) is among the ``last standing'' viable single mediator explanations of both charged and neutral current ``$B$-meson decay anomalies''.
The couplings of such a (gauge) field to SM fermions can be cast as
\begin{equation}\label{eq:modelind:L:massbasis}
	\mathcal L \supset \sum_{i,j,k = 1}^{3} V_1^\mu\left(\bar d_L^i \,\gamma_\mu \,K_L^{ik}\, \ell_L^{k} + \bar u_L^j \,V_{ji}^\dagger\, \gamma_\mu\, K_L^{ik} \,U_{kj}^\mathrm{P}\, \nu_L^j\right) + \mathrm{H.c.}\:\text,
\end{equation}
in which $K_L^{ij}$ are the (effective) LQ couplings to left-handed SM fermions\footnote{For simplicity, and due to the absence of strong hints in the data suggesting otherwise, we restrict the couplings to be only left-handed and real.}, $V$ denotes the Cabibbo-Kobayashi-Maskawa (CKM) quark mixing matrix and $U^\mathrm{P} \equiv U_L^{\ell \dagger} U_L^{\nu}$ is the Pontecorvo-Maki-Nakagawa-Sakata (PMNS) leptonic mixing matrix.
The leading neutral and charged current Wilson coefficients which generate sizeable contributions to $b\to s\ell\ell$ and $b\to c\ell\nu$ can then be conveniently expressed as
\begin{eqnarray}
C^{ij;\ell \ell^{\prime}}_{9,10} &=& \mp\frac{\pi}{\sqrt{2}G_F\,\alpha_\text{em}\,V_{3j}\,V_{3i}^{\ast} \,m_{V_1}^2}\left(K_L^{i
  \ell^\prime} \,K_L^{j\ell\ast} \right)\,,\nonumber \\
  C_{jk,\ell i}^{V_L} &=& \frac{\sqrt{2}}{4\,G_{F}\,m_{V_1}^2}\,
  \frac{1}{V_{jk}}\,
  (V\,K_{L}\, U^P)_{ji}\, K_{L}^{k\ell\ast}\,\text.
  \label{eqn:CV}
\end{eqnarray}
Remarkably, $V_1$ closely connects the charged current with the neutral current anomalies; while an enhancement in $b\to c\tau\nu$ transitions requires sizeable $b\tau$ and $c\nu_\tau$ couplings, the same couplings also generate sizeable $b\to s\tau\tau$ operators which feed under renormalisation group (RG) running (from the LQ mass scale to the observable scale) into $b\to s\mu\mu$ and $b\to see$ operators, as pointed out in~\cite{Crivellin:2018yvo}.
While this behaviour naturally occurs in $V_1$ models attempting at a combined explanation, it is also preferred by the data on $b\to s\ell\ell$ processes on its own.
In the left plot of Figure~\ref{fig:plots1} we present a model independent fit under the hypothesis of (lepton flavour) {\it universal} NP contributions to $\Delta C_{9}^{bs\ell\ell} = \Delta C_9^\text{univ.}$ in addition to the ``$(V-A)$-current'' $\Delta C_9^{bs\mu\mu} = -\Delta C_{10}^{bs\mu\mu}$.
To evaluate the impact of the most recent data on the angular observables in $B\to K^\ast \mu\mu$ decays provided by the LHCb collaboration~\cite{Aaij:2020nrf}, we first establish a baseline by fitting the described NP hypothesis to all available data, excluding the latest measurements.
In a second fit we then include the most recent data into the likelihood. 
In the left panel of Figure~\ref{fig:plots1} we present both fits in the plane of the $(V-A)$ NP contributions vs. the additional {\it universal} contribution in $C_9$. 
While the best-fit point is only slightly affected, the new data-improved precision has a strong impact on the likelihood contour levels.
It can be seen that a non-vanishing {\it universal} contribution to $C_9$ is now preferred by around $3\:\sigma$ compared to formerly $2\:\sigma$.
The resulting correlation between charged and neutral current transitions can further be visually verified in Figure~\ref{fig:plots1} where we show in the right plot a fit of the most dominant LQ coupling combinations relevant to to $b\to s \ell\ell$ and $b\to c\tau\nu$ data.

\section{Non-unitary couplings}
Being a gauge boson\footnote{Vector LQs can also appear as composite fields; for simplicity we here assume it to arise from an extension of $SU(3)_c$.}, the couplings of $V_1$ to SM fermions are a priori universal.
The universality is however in general broken by a possible mass misalignment of quarks and leptons, such that, analogous to the CKM mechanism, a {\it unitary} flavour structure is generated.
In order to accommodate the anomalous data in $B$-meson decays, the LQ couplings $K_L^{ij}$ are required to exhibit a particular flavour pattern, which in turn leads to sizeable charged lepton flavour violation (cLFV) rates.
As it turns out, the experimental bounds on cLFV processes can be respected if the ($3\times3$) coupling matrix $K_L^{ij}$ is {\it non-unitary}~\cite{Hati:2019ufv}.
Non-unitary couplings (to SM fermions) can be achieved by mixings of SM fermions with additional {\it vector-like} fermions. 
Inspired from neutrino physics~\cite{Xing:2007zj}, the effects of these mixings, specifically with leptons, can be parametrised in an effective way, using a product of unitary rotation matrices.
Should a specific model encompass additional vector-like leptons, one must also consider the impact of the new states for electroweak precision observables, as for instance the constraints on the $Z$ boson LFU ratios and cLFV decay modes.
In particular, only vector-like $SU(2)_L$ doublets allow to circumvent otherwise excessive corrections to the $Z-\ell\ell^{(\prime)}$ tree-level couplings, as emphasised in~\cite{Hati:2019ufv}.

\section{Results and Outlook}
\begin{figure}
    \centering
    \mbox{
    \hspace{-3mm}
    \includegraphics[width=0.5\textwidth]{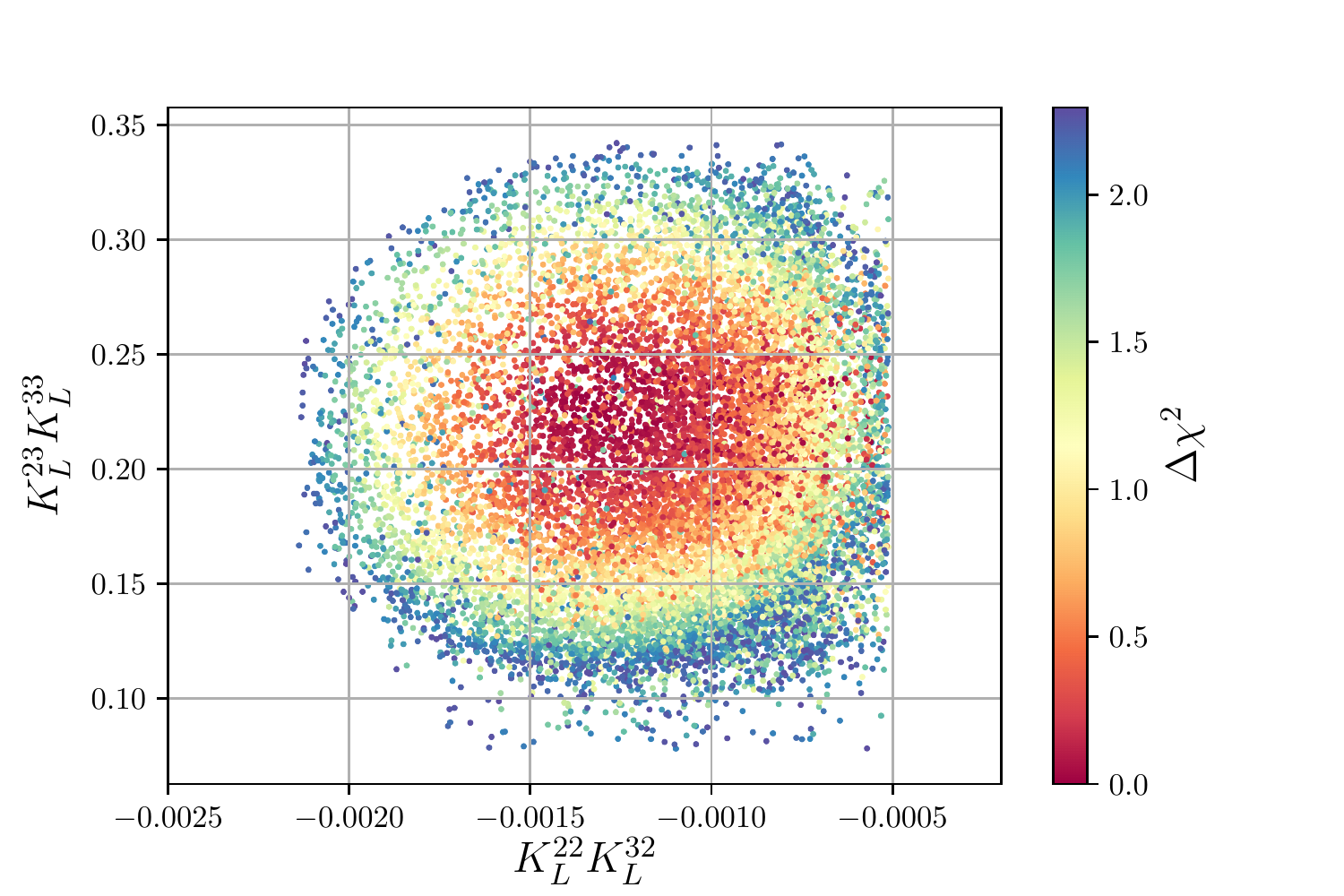}
    \raisebox{-6mm}{\includegraphics[width=0.55\textwidth]{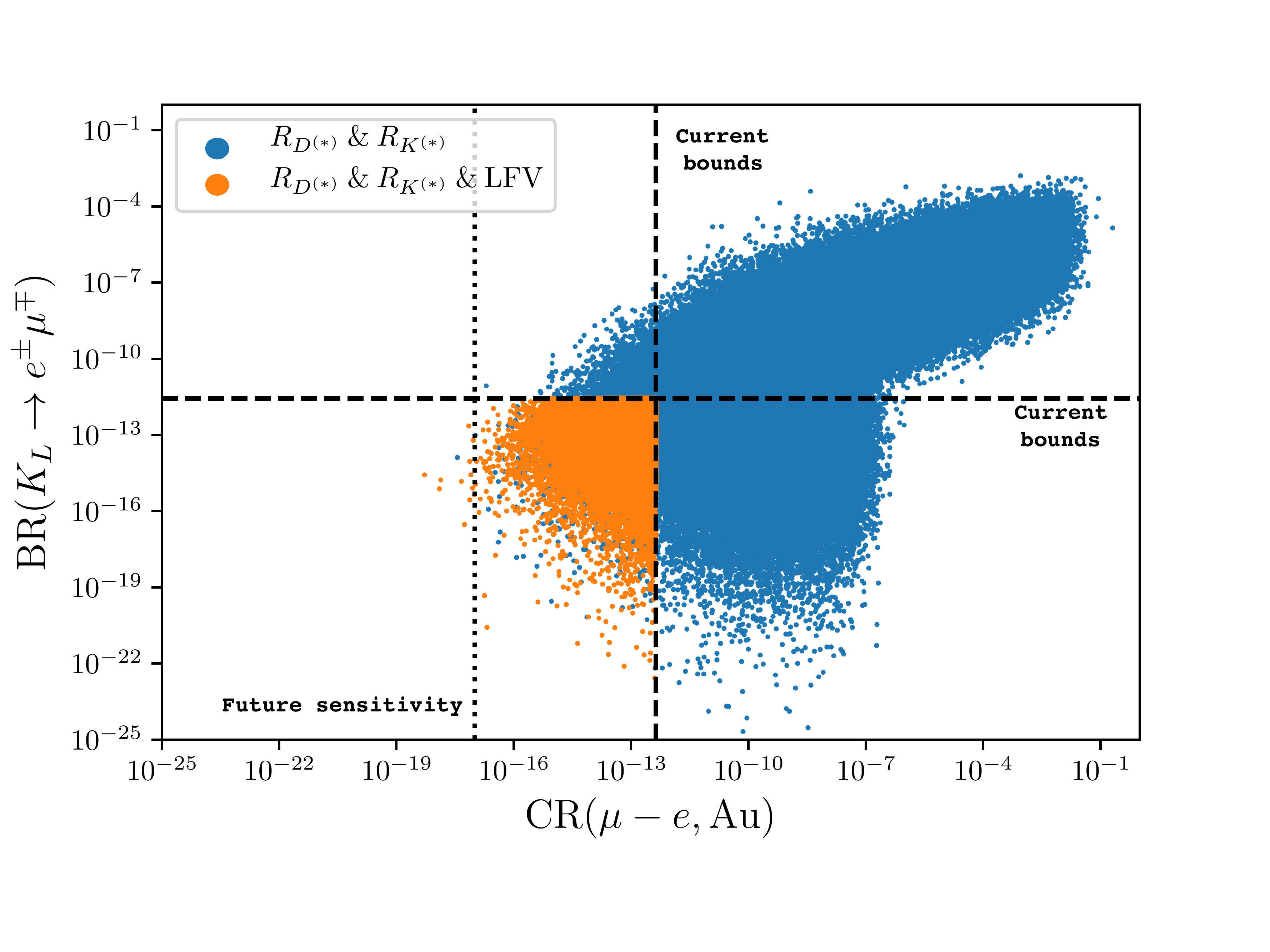}}
    }
    \vspace*{-7mm}
    \caption{{\bf Left:} Sample points reconciling the $B$-meson decay anomalies at $1\:\sigma$ while being consistent with current experimental bounds on cLFV processes.
    {\bf Right:} Sample points (of the model) in the plane of the most constraining cLFV observables. Points in blue accommodate both anomalies at the $3\:\sigma$-level, points in orange are in addition complying with all cLFV bounds. Dashed lines denote current experimental bounds, while the dotted line indicates the future sensitivity of COMET~\cite{Adamov:2018vin} and Mu2e~\cite{Bartoszek:2014mya}.}
    \label{fig:plots2}
\end{figure}

In a specific construction\footnote{Contrary to many other attempts in the literature, we do not neglect any coupling in order to comply with cLFV bounds; {\it all} couplings are a priori assumed to be non-vanishing.} including 3 generations of vector-like doublet leptons, we explored the model's parameter space by randomly scanning over 12 (of 15 total) relevant mixing angles at a mass benchmark point of $m_{V_1} = 1.5\:\mathrm{TeV}$. 
Our phenomenological analysis includes, besides a large set of $b\to s\ell\ell$ and $b \to c\ell\nu$ observables, numerous constraints from cLFV decay modes of light and heavy mesons as well as purely leptonic cLFV processes.
The results of this scan are shown in the left panel of Figure~\ref{fig:plots2}, where we present a large number of sample points in the plane of the dominant combinations of LQ couplings: these allow reconciling the anomalous data at the $1\:\sigma$ level, agreeing with the parametrisation independent fit shown in the right panel of Figure~\ref{fig:plots1}. 
Although a significant number of additional parameters is introduced, solid predictions can be made.
In the right panel of Figure~\ref{fig:plots2} we show the sample points in the plane of the most constraining observables $K_L\to \mu^\pm e^\mp$ and neutrinoless $\mu - e$ conversion.
It can be seen that a large portion of the parameter space is already excluded (blue points), while the remaining (preferred) region (orange points) is well within the reach of the upcoming cLFV-dedicated experiments COMET~\cite{Adamov:2018vin} and Mu2e~\cite{Bartoszek:2014mya}.

Besides the excellent experimental prospects from cLFV-dedicated experiments (COMET/Mu2e and Mu3e~\cite{Blondel:2013ia}), Belle II recently started its first data-taking run.
Being a $B$-factory, it aims not only to scrutinise the $B$-anomalies, but also to produce copious amounts of charmed mesons and $\tau$ leptons, allowing to reach an unprecedented precision and sensitivity on related flavour processes in the near future.
Going beyond a random scan, and to a comprehensive statistical analysis (a global fit of the LQ couplings) allows to explore the experimental prospects on how vector LQ models can be further probed in the near future~\cite{LQ2020}.
By means of Monte-Carlo techniques we thus estimate ranges for numerous observables to be investigated by Belle II~\cite{Kou:2018nap}, for three LQ mass benchmark points $m_{V_1}\in \{1.5, 2.5, 3.5 \}\:\mathrm{TeV}$.
\begin{figure}
    \centering
    \includegraphics[width=\textwidth]{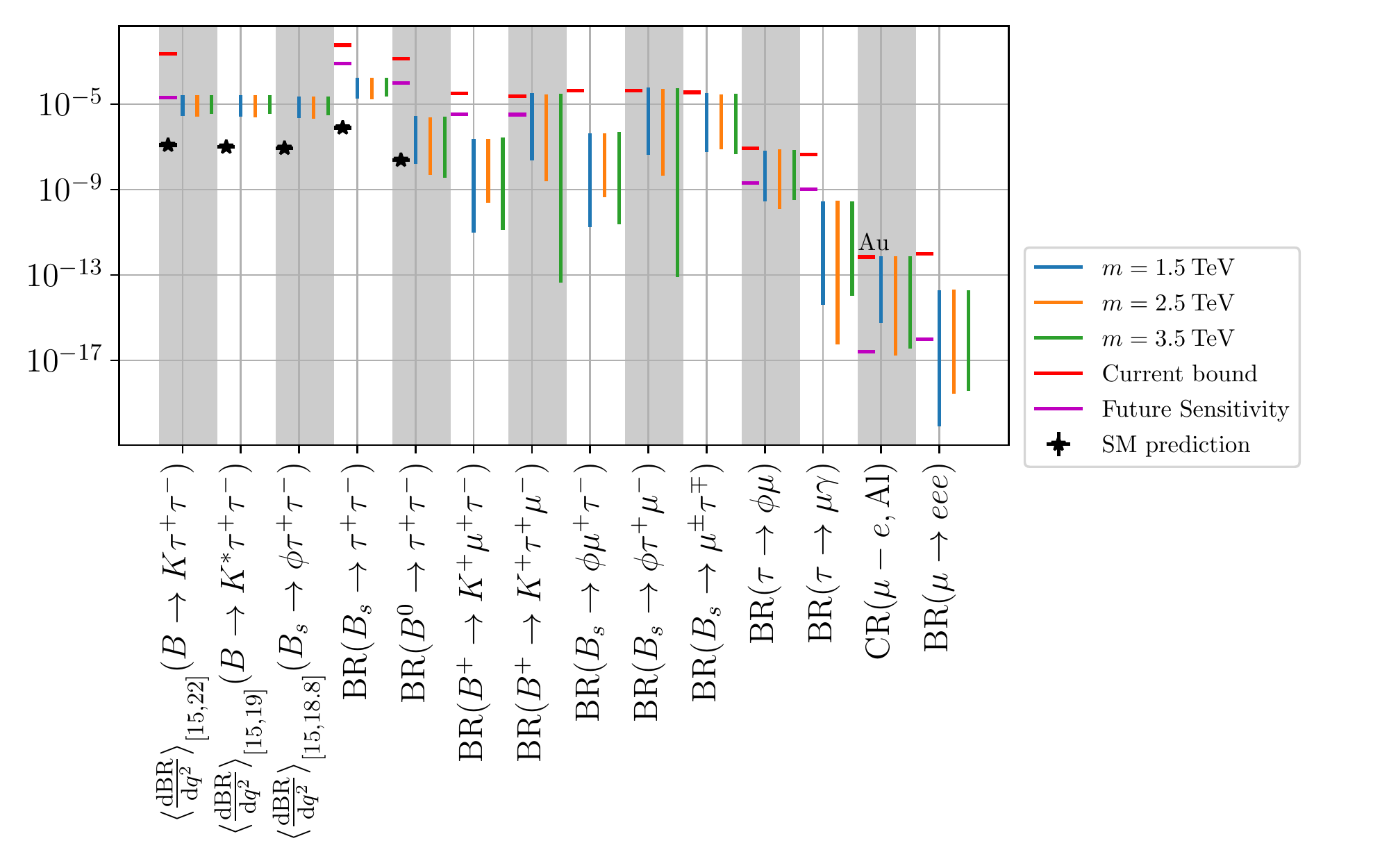}
    \caption{Predicted ranges for several $b\to s\tau\tau$ and cLFV observables to be investigated at Belle II~\cite{Kou:2018nap}, COMET~\cite{Adamov:2018vin}/Mu2e~\cite{Bartoszek:2014mya} and Mu3e~\cite{Blondel:2013ia}. The blue, orange and green lines respectively denote the $90\%$-range around the respective best-fit points, for leptoquark masses of $1.5,\, 2.5\text{ and } 3.5\:\mathrm{TeV}$. The horizontal red (purple) lines denote the current (future) bound at $90\,\%$ C.L.;  stars denote SM predictions when appropriate. All Belle II sensitivities correspond to an integrated luminosity of $50\:\mathrm{ab}^{-1}$.}
    \label{fig:taulfv}
\end{figure}
As previously mentioned, due to the $SU(2)_L$ representation of $V_1$, the LQ couplings responsible for sizeable NP contributions in $b\to c\tau\nu$ operators, as required by $R_{D^{(\ast)}}$ data, also generate large NP contributions in $b\to s\tau\tau$ operators.
Consequently, significant enhancements in related meson decays, with respect to the SM prediction, are expected.
Additionally, a successful accommodation of the anomalous $b\to s\ell\ell$ data necessitates large $b\mu$ and $s\mu$ LQ couplings.
This combination induces $\tau-\mu$ flavour violating processes at the tree-level, in many cases at testable rates.
In Figure~\ref{fig:taulfv} we present (preliminary) predictions for several $b\to s\tau\tau$ and $\tau-\mu$ flavour violating observables together with current experimental bounds and future sensitivities. 
In addition to neutrinoless $\mu - e$ conversion in nuclei, $b\to s\tau\tau$ and $b \to s\tau\mu$ related meson decays, as well as the cLFV decay $\tau\to\phi\mu$ can be singled out as ``{\it golden modes}''.
Especially $\tau\to\phi\mu$ decays have excellent chances of being observed by Belle II. 
Conversely, should future searches for the decays yield negative results, the related LQ couplings would be tightly constrained. It then might prove very challenging to accommodate the $B$-meson decay anomalies with a single vector LQ.

\section{Summary and Conclusion}
In recent years, numerous observables hint at the presence of LFUV in charged and neutral current semileptonic $B$-meson decays.
$SU(2)_L$-singlet vector leptoquarks are excellent candidates for a combined explanation, but are subject to stringent constraints from cLFV observables.
We have shown that, in order to comply with data on a large number of flavour observables, the flavour structure of the LQ couplings to SM matter are required to be non-unitary, which can be generated by mixings of SM leptons with heavy vector-like doublet states.
Furthermore, we have made predictions for many $b\to s\tau\tau$ and cLFV observables, out of which we have identified several ``{\it golden modes}'', that have excellent chances to be observed by upcoming experiments in the near future.

\end{document}